\documentclass{ws-procs10x7}
\usepackage{balance}

\newcolumntype{d}[1]{D{.}{.}{#1}}

\makeindex
\begin{document}

\title{Search for direct CP-violation in $K^\pm\rightarrow 3\pi$ decays by NA48/2}

\author{S. BALEV$^*$}

\address{Laboratory of Particle Physics, Joint Institute for Nuclear Research, Dubna,  141980, Russia\\$^*$E-mail: balev@sunse.jinr.ru}

\twocolumn[\maketitle\abstract{A high precision search for direct
CP-violation in $K^\pm\rightarrow 3\pi$ decays was performed by the
NA48/2 experiment at CERN SPS. The asymmetry in the Dalitz plot
linear slopes $A_g=(g^+-g^-)/(g^++g^-)$ is measured to be
$A^c_g=(-1.3\pm2.3)\cdot10^{-4}$ by studying $\sim3.1\cdot10^9$
$K^{\pm}\rightarrow\pi^\pm\pi^+\pi^-$ decays and
$A^n_g=(2.1\pm1.9)\cdot10^{-4}$ by studying $\sim91\cdot10^6$
$K^{\pm}\rightarrow\pi^\pm\pi^0\pi^0$ decays. The unique double-beam
system, the design of the detectors and the method of analysis
provide good control of the instrumental charge asymmetries and
allow to keep the precision of the result limited by statistics,
reaching accuracy one order of magnitude better than in previous
experiments.} \keywords{CP-violation, Kaon physics.} ]

\section{Introduction}
More than 40 years after its discovery in the mixing of neutral
kaons, the full understanding of CP violation is far from being
reached. Two recent breakthroughs spread light over this puzzling
phenomenon: In the late 90s the experiments NA48 and KTeV confirmed
the earlier indication from NA31 experiment for direct CP-violation
in $K^0$ system. Secondly, in the beginning of this century the
CP-violating processes were found in B mesons field by the
experiments Belle and Babar. In order to explore possible
non-Standard Model (SM) enhancements to heavy-quark loops which are
at the core of direct CP-violating processes, various systems must
be studied. In kaons, besides the $\epsilon^\prime/\epsilon$
parameter in $K^0\rightarrow\pi\pi$ decays, promising complementary
observables are the rates of GIM-suppressed rare kaon decays
proceeding through neutral currents, and an asymmetry between $K^+$
and $K^-$ decays to three pions.

The $K^\pm\rightarrow 3\pi$ matrix element can be parameterized by a
polynomial expansion in two Lorentz-invariant variables $u$ and $v$:
\begin{equation}
\left|M\left(u,v\right)\right|^2\propto1+gu+hu^2+kv^2+...,
\end{equation}
where $|h|, |k|\ll|g|$ are the slope parameters, and
\begin{equation}
u=\frac{s_3-s_0}{m^2_\pi},\quad{} v=\frac{s_1-s_2}{m^2_\pi},
\end{equation}
where $m_\pi$ is the charged pion mass,
$s_i=\left(p_K-p_i\right)^2$, $s_0=\sum s_i/3$ $(i=1,2,3)$, $p_K$
and $p_i$ are kaon and $i$-th pion 4-momenta, respectively. The
index $i=3$ corresponds to the odd pion\footnote{The other two pions
have the same charge.}. The parameter of direct CP violation is
usually defined as:
\begin{equation}
A_g=\frac{g^+-g^-}{g^++g^-},
\end{equation}
where $g^+$ is the linear coefficient in (1) for $K^+$ and $g^-$ --
for $K^-$. A deviation of $A_g$ from zero is a clear indication for
direct CP violation. Several experiments~\cite{6} have searched for
such asymmetries both in $K^\pm\rightarrow\pi^\pm\pi^+\pi^-$ and
$K^\pm\rightarrow\pi^\pm\pi^0\pi^0$ decay modes, and obtained
consistent with zero result with a precision at the level of
$10^{-3}$. SM predictions for $A_g$ vary from a few $10^{-6}$ to a
few $10^{-5}$~\cite{5}, however some theoretical calculations
involving processes beyond the SM~\cite{7} predict substantial
enhancements of the asymmetry, which could be observed in the
present experiment.

\section{Description of the Experiment}

\begin{figure*}[t]
\begin{center}
\resizebox{.9\textwidth}{!}{\includegraphics{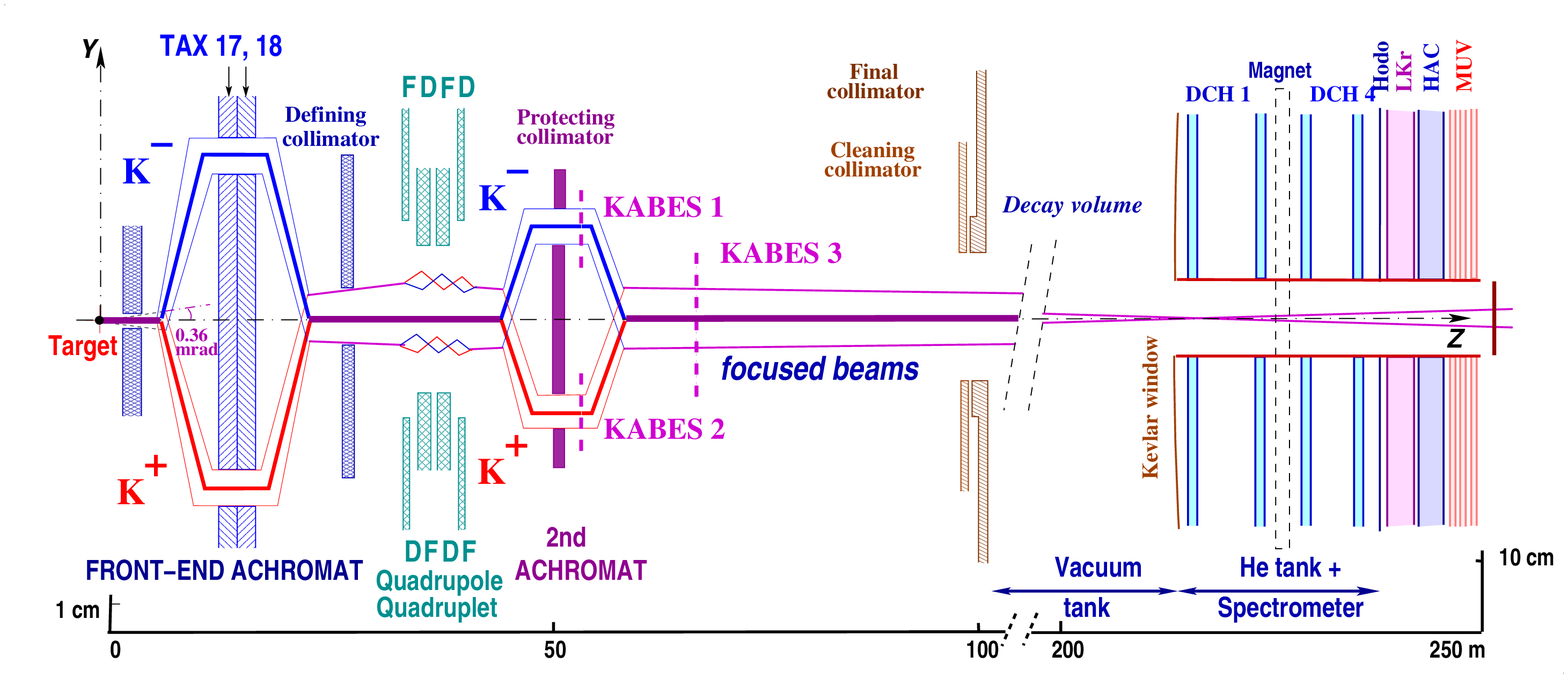}}
\end{center}
\vspace{-6mm} \caption{Schematic lateral view of the NA48/2
experiment. Region 1 (from target to decay volume): beam line
(TAX17,18: motorized beam dump/collimators used to select the
momentum of the $K^+$ and $K^-$ beams; DFDF: focusing quadrupoles;
KABES1-3: beam spectrometer stations). Region 2: decay volume and
detector (DCH1-4: drift chambers; Hodo: hodoscope; LKr:
electromagnetic calorimeter; HAC: hadron calorimeter; MUV: muon
veto). The vertical scales for the two regions are different.}
\label{beamline}
\end{figure*}

The NA48/2 experiment at the CERN SPS was designed to search for
direct CP-violation in the decays of charged kaons to three pions,
and collected data in 2003 and 2004. In order to reach a high
accuracy in the measurement of the charge asymmetry parameter $A_g$,
the highest possible level of charge symmetry between $K^+$ and
$K^-$  was a crucial requirement in the choice of beam, experimental
apparatus, strategy of data taking and analysis.

A novel beam line (Fig.~\ref{beamline}) with two simultaneous
charged beams of opposite charges was designed and built in the high
intensity hall (ECN3) at the CERN SPS. The charged particle beams
were produced by 400 GeV protons with high intensity from the SPS
impinging on a beryllium target. Charged particles with momentum
$(60\pm3)$ GeV/$c$ were selected symmetrically by an achromatic
magnet system ('achromat') which separates vertically the two beams
and recombines them again on the same axis. Frequent inversion of
the magnetic field polarities in all the beamline elements provides
a high level of intrinsic cancellation of the possible systematic
effects in the beam line. The $K^+/K^-$ flux ratio is about 1.8.

The entire reconstruction of $K^\pm\rightarrow\pi^\pm\pi^+\pi^-$
decays and the determination of the kaon charge in
$K^\pm\rightarrow\pi^\pm\pi^0\pi^0$ decays rely on a magnetic
spectrometer. Two drift chambers are located upstream and two
downstream of a dipole magnet which deflects charged particles
horizontally with a transverse momentum kick of 120 MeV/$c$. The
magnetic field was reversed frequently in order to cancel possible
left-right asymmetries in the detector system. The momentum
resolution of the magnetic spectrometer is
$\sigma(p)/p=1.0\%\oplus0.044\%p$ ($p$ in GeV/$c$). The acceptance
of the spectrometer is defined mainly by an evacuated beam tube
passing through its centre, with a diameter of $\sim16$ cm, in which
the surviving beam particles as well as the muons from
$\pi\rightarrow\mu\nu$ decays travel.

The reconstruction of $K^\pm\rightarrow\pi^\pm\pi^0\pi^0$ decays is
based mainly on the use of liquid krypton calorimeter (LKr), which
measures the energies of the four photons from $\pi^0$ decays. The
LKr has an energy resolution
$\sigma(E)/E=0.032/\sqrt{E}\oplus0.09/E\oplus0.0042$ ($E$ in GeV)
and spatial resolution for a single electromagnetic shower
$\sigma_x=\sigma_y=0.42/\sqrt{E}\oplus0.06$ cm for the transverse
coordinates $x$ and $y$. The use of \`{a} priori charge symmetric
detector helps to keep the result in
$K^\pm\rightarrow\pi^\pm\pi^0\pi^0$ mode practically unbiased.

A hodoscope is used for precise time measurement of the charged
particles and as a component in the trigger system for both decay
modes. Detailed description of the detector components can be found
elsewhere~\cite{12}.

\section{Asymmetry measurement method}
The asymmetry measurement is based on the comparison of the $u$
spectra for $K^+$ and $K^-$ decays, $N^+(u)$ and $N^-(u)$,
respectively. The ratio of the $u$ spectra $N^+(u)/N^-(u)$ is
proportional to
\begin{equation}
\frac{N^+(u)}{N^-(u)}\propto\frac{1+(g+\Delta g)u+hu^2}{1+gu+hu^2},
\end{equation}
where $g$ and $h$ are the actual of the Dalitz-slope
parameters~\cite{13}. The possible presence of a direct CP violating
difference between the linear slopes of $K^+$ and $K^-$, $\Delta
g=g^+-g^-$, can be extracted from a fit to this ratio. The measured
asymmetry is then given by $A_g=\Delta g/2g$.

In order to minimize the effect of beam and detector asymmetries, we
use the ratio $R_4(u)$, defined as a product of four $N^+(u)/N^-(u)$
ratios:
\begin{eqnarray}
R_4(u) & = & R_{US}\cdot R_{UJ}\cdot R_{DS}\cdot R_{DJ}=\nonumber
\\[4pt]
&&{} = R\left(1+\frac{\Delta g\cdot u}{1+gu+hu^2}\right)^4.
\end{eqnarray}
The first subscript $U$ ($D$) denots a configuration of the beam
magnet polarities which corresponds to the positive beam traversing
the upper (lower) path in the achromats. The second subscript $S$
denotes the spectrometer magnet polarities (opposite for the events
in the numerator and in the denominator of each ratio) deflecting
the kaon to negative $x$ (towards the Sal\`{e}ve mountain, given the
topographical situation of the experiment in relation to the
mountains surrounding CERN) and $J$ corresponds to the deflection of
the pions in the opposite direction (towards the Jura mountain
chain). The spectra $N^+(u)$ and $N^-(u)$ for each of the four
individual ratios in (5) are obtained from successive runs taken
with the same beam magnet polarities and with the $\pi^{\pm}$
deflected in the same direction by the spectrometer magnet. The
parameter $\Delta g$ and the normalization $R$ are extracted from a
fit to the measured quadruple ratio $R_4(u)$ using the function in
eq. (5)\footnote{Due to smallness of the parameters $g$ ad $h$ in
$K^{\pm}\rightarrow\pi^\pm\pi^+\pi^-$ mode, a fit with a function
$R_4(u)=R(1+\Delta g\cdot u)^4$ is sufficient.}. The measured slope
difference $\Delta g$ is insensitive to the normalization parameter
$R$, which reflects the ratio of $K^+$ and $K^-$ fluxes.

The quadruple ratio method complements the procedure of magnet
polarity reversal. It allows a three-fold cancellation of systematic
biases: 1) beam line biases cancel between $K^+$ and $K^-$ samples
in which the beams follow the same path; 2) the effect of local
non-uniformities of the detector cancel between $K^+$ and $K^-$
samples in which charged pions illuminate the same parts of the
detectors; 3) as a consequence of using simultaneous $K^+$ and $K^-$
beams, global, time dependent, instrumental charge asymmetries
cancel between $K^+$ and $K^-$ samples.

A reduction of possible systematic biases due to the presence of
stray permanent magnetic fields (Earth field, vacuum tank
magnetization) is achieved by the radial cuts around the average
beam position, which make the geometrical acceptance to charged
pions azimuthally symmetric. The only residual sensitivity to
instrumental charge asymmetries is associated with time variations
of any acceptance asymmetries occurring on a time scale shorter than
the magnetic field alternation period, which are studied carefully
by a number of monitors recorded throughout data taking.

\section{Result and conclusions}
The whole 2003 and 2004 data-set contains several\footnote{9 for
$K^{\pm}\rightarrow\pi^\pm\pi^+\pi^-$ and 7 for
$K^{\pm}\rightarrow\pi^\pm\pi^0\pi^0$ decay mode.} samples with all
possible combinations of magnetic field polarities in the beam
optics and in the magnetic spectrometer, needed to construct the
quadruple ratio (5) and self-sufficient for $\Delta g$ measurement.
The raw asymmetry is extracted for each of them separately, and the
final result then is obtained as their weighted average. In total,
$3.1\cdot10^9$ $K^{\pm}\rightarrow\pi^\pm\pi^+\pi^-$ and
$91\cdot10^6$ $K^{\pm}\rightarrow\pi^\pm\pi^0\pi^0$ decays were
selected for the analysis. The result in terms of linear slope
difference $\Delta g$ with only the statistical error quoted is
$\Delta g^c=(0.7\pm 0.7)\cdot10^{-4}$ for
$K^{\pm}\rightarrow\pi^\pm\pi^+\pi^-$ and $\Delta g^n=(2.7\pm
2.0)\cdot10^{-4}$ for $K^{\pm}\rightarrow\pi^\pm\pi^0\pi^0$ decay
mode. These results are free of systematic biases in the first
approximation due to the implemented method of cancellation of
various apparatus imperfections. However, the checks of possible
systematic contributions have been done, and corresponding
uncertainties were obtained~\cite{asym}. The preliminary results for
the whole 2003 and 2004 data-set are:
\begin{equation}
A_g^c = (-1.3\pm 1.5_{stat.}\pm 0.9_{trig.}\pm 1.4_{syst.})\cdot
10^{-4}
\end{equation}
\begin{equation}
A_g^n = (2.1\pm 1.6_{stat.}\pm 1.0_{syst.}\pm 0.2_{ext.})\cdot
10^{-4}
\end{equation}
correspondingly for $K^{\pm}\rightarrow\pi^\pm\pi^+\pi^-$ and
$K^{\pm}\rightarrow\pi^\pm\pi^0\pi^0$ decay modes\footnote{The
trigger error 0.9 in (6) is an upper limit for eventual charge
asymmetric response of the trigger system for
$K^{\pm}\rightarrow\pi^\pm\pi^+\pi^-$ decays and it is limited by
the statistics in the control sample used for this estimation; for
$K^{\pm}\rightarrow\pi^\pm\pi^0\pi^0$ decays this error is included
in the total systematic error. The external uncertainty 0.2 in (7)
arises from the experimental precision on $g$ and $h$~\cite{13}.}.
The reason for a similar precision of the results given in (6) and
(7), despite a factor of $\sim30$ in the collected statistics, is
the fact that the population density of the Dalitz plot is more
favourable in the $K^\pm\rightarrow\pi^\pm\pi^0\pi^0$ mode and
$|g(\pi^\pm\pi^0\pi^0)|\sim 3\cdot|g(\pi^\pm\pi^+\pi^-)|$. The
results are one order of magnitude more precise than previous
measurements and are consistent with the predictions of the SM.

\end{document}